\documentclass[letterpaper,10pt]{article} 

\usepackage{opticameet3} 
\usepackage{listings}
\newcommand\authormark[1]{\textsuperscript{#1}}

\usepackage{amsmath,amssymb}
\usepackage[colorlinks=true,bookmarks=false,citecolor=blue,urlcolor=blue]{hyperref} 

\begin{document}

\title{Assessing the Elephant in the Room in Scheduling for Current Hybrid HPC-QC Clusters}


\author{Paolo Viviani\authormark{1,*} Roberto Rocco\authormark{2}, Matteo Barbieri\authormark{2}, Gabriella Bettonte\authormark{2}, Elisabetta Boella\authormark{2}, Marco Cipollini\authormark{4}, Jonathan Frassineti\authormark{3}, Fulvio Ganz\authormark{2}, Sara Marzella\authormark{3}, Daniele Ottaviani\authormark{3}, Simone Rizzo\authormark{2}, Alberto Scionti\authormark{1}, Chiara Vercellino\authormark{1}, Giacomo Vitali\authormark{1, 4}, Olivier Terzo\authormark{1}, Bartolomeo Montrucchio\authormark{4} and Daniele Gregori\authormark{2}  }

\address{\authormark{1}\textit{LINKS Foundation}, Torino, Italy\\
\authormark{2}\textit{E4 Computer Engineering}, Scandiano, Italy\\
\authormark{3}\textit{CINECA}, Casalecchio di Reno, Italy\\
\authormark{4}\textit{Politecnico di Torino}, Torino, Italy\\
}

\email{\authormark{*}paolo.viviani@linksfoundation.com} 


\begin{abstract}
Quantum computing resources are among the most promising candidates for extending the computational capabilities of High-Performance Computing (HPC) systems. As a result, HPC–quantum integration has become an increasingly active area of research. While much of the existing literature has focused on software stack integration and quantum circuit compilation, key challenges such as hybrid resource allocation and job scheduling—especially relevant in the current Noisy Intermediate-Scale Quantum era—have received less attention. In this work, we highlight these critical issues in the context of integrating quantum computers with operational HPC environments, taking into account the current maturity and heterogeneity of quantum technologies. We then propose a set of conceptual strategies aimed at addressing these challenges and paving the way for practical HPC-QC integration in the near future.
\end{abstract}

\section{Introduction}
Nowadays, Quantum Computers (QCs) are considered an excellent candidate for accelerating computationally expensive tasks in chemistry and materials research. These fields of study also represent typical examples of research domains for High Performance Computing (HPC); therefore, hybrid classical-quantum approaches are desirable and actively researched.
In the context of the ongoing effort to integrate QCs and HPC systems, a clear trend emerges: as technology matures, the programming model is likely to shift toward the well-established offloading approach, similar to what is currently used for Graphical Processing Unit (GPU) accelerators~\cite{Elsharkawy_2024}. 
However, the \emph{current} scenario that features co-existence between different technologies with different characteristics requires a set of specific tools and techniques that must address a critical issue: \emph{Quantum Processing Units} (QPUs) \emph{are} \emph{a scarce resource}.

To enable the integration of real HPC workloads (i.e., large-scale MPI jobs) with quantum jobs beyond the well-known challenges related to compilers, software stack, programming languages, etc.~\cite{beck2024quantum,humbleQuantumComputersHighPerformance2021,giusto_2024}, we claim that addressing the fair and efficient allocation of quantum and classical resources is a crucial step. Previous literature~\cite{schulzAcceleratingHPCQuantum2022a,brittHighPerformanceComputingQuantum2017,Saurabh_middleware_2023}
highlighted the importance of dedicated scheduling strategies—particularly in scenarios where multiple users or jobs share access to a single QPU. Without such strategies, both quantum and classical resources risk being underutilised or wasted. 
With this paper, we aim to further underline these criticalities and propose some ideas to address these issues. In particular:
\begin{itemize}
\item We provide a brief overview of the current state of near-term HPC-QC integration; 
\item We identify some key challenges to achieving integration within operational HPC environments, focusing on resource allocation and scheduling;
\item We propose and critically evaluate a set of solutions aimed at addressing these challenges, highlighting their potential benefits and limitations.
\end{itemize}

The paper is structured as follows: Section~\ref{sec:sota} reviews recent work on HPC-QC integration, highlighting key concepts in quantum resource allocation. Section~\ref{sec:challenges} outlines specific integration challenges, while Section~\ref{sec:approach} proposes approaches to address them within operational HPC constraints. Section~\ref{sec:conclusion} concludes the paper. 


\section{HPC-QC integration overview}\label{sec:sota}

HPC-QC integration is actively investigated at the research level~\cite{schulzAcceleratingHPCQuantum2022a,humbleQuantumComputersHighPerformance2021,bartschvaleriaQCHPCQuantum2021,brittHighPerformanceComputingQuantum2017,mccaskeyXACCSystemlevelSoftware2020}, promoted by quantum computing vendors~\cite{ruefenachtBringingQuantumAcceleration}, and supported by prominent initiatives such as EuroHPC\footnote{\url{https://doi.org/10.3030/101018180}} and the Quantum Flagship framework \cite{binosiEuroQCSEuropeanQuantum}. These efforts contribute significantly to the classification of quantum applications~\cite{humbleQuantumComputersHighPerformance2021}, the definition of HPC-QC architectures~\cite{bartschvaleriaQCHPCQuantum2021,humbleQuantumComputersHighPerformance2021,beck2024quantum}, the development of a shared software stack~\cite{humbleQuantumComputersHighPerformance2021,schulzAcceleratingHPCQuantum2022a,willeMQT}, and the initial outline of programming models~\cite{mccaskeyXACCSystemlevelSoftware2020, schulzAcceleratingHPCQuantum2022a}. 

Currently, however, quantum machines are typically installed in lab environments, far from operational data centres, and accessed via a public network through high-level languages and dedicated libraries (e.g., Qiskit, Pulser). Through these tools, users can define kernels (i.e., circuits), send them to the machine via Representational State Transfer (REST) Application Programming Interfaces (APIs), and execute them in a sequential, mostly single-threaded fashion.


While the cloud access model is effective for utilising current quantum machines, it falls short of achieving true HPC integration, which demands direct, high-bandwidth connectivity and closed-loop control over the quantum hardware (HW). The research community, including HPC facilities, has begun exploring steps in this direction \cite{Esposito_2023,Elsharkawy_2024,shehata_2025,mantha_2024,beck2024quantum}, but there is no clear consensus on how HPC-QC integration should be realised, as the technology is still in its early stages. Nevertheless, we can categorise HPC-QC integration approaches into three types, based on the solutions proposed in the literature~\cite{humbleQuantumComputersHighPerformance2021,bartschvaleriaQCHPCQuantum2021,Saurabh_middleware_2023}:
\begin{itemize}
    \item \emph{Tightly coupled}: classical computation happens within the coherence time of the quantum machine
    , and affects the execution of the circuit. The main use case for this is error correction (i.e., by mid-circuit measurements, syndrome extraction, and decoding), which currently represents more a quantum HW challenge than a matter of HPC-Quantum integration. 
    \item \emph{Quantum offloading}: typical circuit execution is used to accelerate specific operations (i.e., quantum kernels). It can be either synchronous or asynchronous, but the classical code manages the control flow and continues running until completion.
    \item \emph{Loosely-coupled} (workflows): an external workflow manager (even as simple as a shell script) handles the control flow. Classical and quantum resources are allocated separately for each step once that step is ready to run.
\end{itemize}
The first mode still requires a significant advancement in HW maturity, hence it will not be discussed in the scope of this paper, which focuses on the use-cases for the second and third integration modes.

\section{Current HPC-QC integration challenges}\label{sec:challenges}
\begin{figure}[t]
  \centering
  \includegraphics[width=0.9\linewidth]{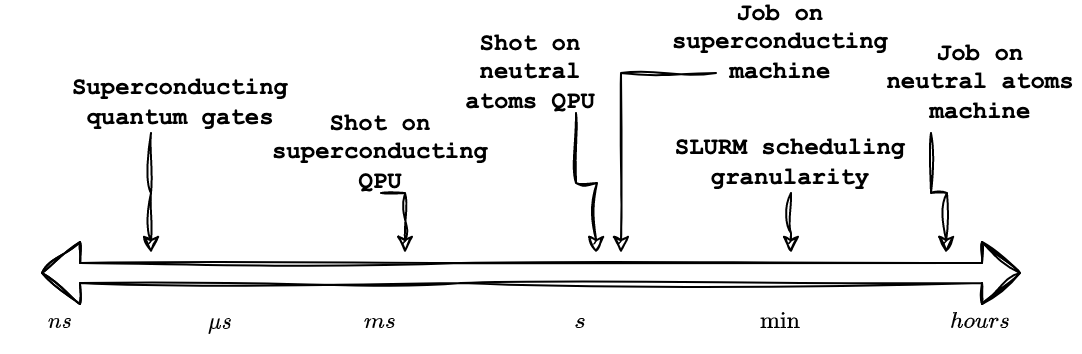}
  \caption{Time scales of relevant quantum jobs/shots. Jobs on neutral atoms machines include the calibration time for an arbitrary register geometry.}\label{fig:timescale}
\end{figure}
When exploring the integration of HPC and quantum computing, it is essential to bridge the fundamental gap between mature, standardised HPC technologies and the still largely experimental nature of quantum HW. In this section, we highlight two key differences that must be addressed to enable effective integration of HPC and quantum resources. We then discuss the challenges of managing access to resources with highly imbalanced availability.

\noindent\textbf{Software heterogeneity}:
The software ecosystems of QC and HPC differ significantly, starting with their choice of programming languages. QC applications often rely on interpreted languages such as Python and Julia, whereas HPC traditionally uses compiled, performance-oriented languages like C, C++, and Fortran, along with programming models such as MPI, OpenMP, and CUDA. This divergence reflects the maturity gap between the two fields: QC is still in its early stages, with a focus on demonstrating usability rather than maximising performance, while HPC is built around mature architectures optimised for efficiency. Although aligning and standardising the software stacks of QC and HPC is essential for future integration, it lies beyond the scope of this work.



\noindent\textbf{Access and allocation model}:
Current quantum computers are typically accessed via dedicated libraries and REST APIs, supported by internal queuing systems that enable asynchronous job submission and multi-user usage. Access to these APIs is generally managed through proprietary accounting mechanisms. However, this model does not align with operational HPC environments, where resource access is governed by job schedulers and policy-driven allocation. Additionally, each quantum HW vendor provides its own API, requiring custom integration efforts for interoperability with existing HPC systems.
Given this context, the first critical challenge is integrating QPU vendor mechanisms into environments governed by batch schedulers and well-established policies and accounting systems. Specifically, we need a way to incorporate QPUs, often exposed as web services, into the operational model of HPC systems, ensuring compatibility with job schedulers like SLURM \cite{slurm} and adherence to institutional resource management policies.

\noindent\textbf{Workload imbalance}:
\begin{lstlisting}[basicstyle=\ttfamily\footnotesize,float,floatplacement=t,language=Bash, caption=Example of SLURM job script to run a potential HPC-QC hybrid job, label=lst:hetjob]
#!/bin/bash
#SBATCH --partition classical
#SBATCH --nodes 10
#SBATCH --time=01:00:00
#SBATCH hetjob
#SBATCH --partition quantum
#SBATCH --gres=qpu:1
#SBATCH --time=01:00:00

srun ./hybrid_job
\end{lstlisting}
Beyond the access model, it is necessary to understand that the characteristic time scales of quantum jobs can heavily affect the job scheduling. Unlike classical computations, which scale with algorithmic complexity and input size, quantum kernel durations are largely dictated by the specific QPU technology. 
As shown in Fig.~\ref{fig:timescale}, execution times can vary significantly depending on the quantum HW, potentially leading to workload imbalances. This variability can challenge the effectiveness of traditional HPC batch schedulers, which may struggle to maintain efficient resource utilization in such heterogeneous environments.

To clarify this point, we can use as an example the job in Listing~\ref{lst:hetjob}, describing a job \emph{co-scheduling} 10 nodes in a classical partition and 1 QPU (defined as a SLURM gres, more on that in Section~\ref{sec:approach}) in a quantum partition, both for 1 hour.
Let us consider a superconducting QPU for which we can assume that each quantum task will last $\sim10 \, \textrm{s}$ (see Fig.~\ref{fig:timescale}): even with many tasks launched inside the job, this can lead to a heavy under-utilisation of the QPU that is exclusively allocated to this job for one hour. Conversely, if the QPU is based on neutral atoms, a quantum task could easily last more than $30 \, \textrm{min}$, leading to under-utilisation of the classical nodes that will be idle waiting for the quantum job completion. These issues must be considered when integrating quantum HW into HPC facilities to ensure efficient resource allocation: \emph{simple co-scheduling with exclusive QPU access is inadequate for achieving optimal resource utilization in heterogeneous HPC-QC environments}. 

\section{Proposed integration ideas}\label{sec:approach}



Given the current heterogeneity in QC and hybrid algorithms, a one-size-fits-all solution is unlikely. Instead, we envision a small set of HPC-friendly, complementary approaches tailored to address integration challenges based on the direction of workload imbalance. While these strategies are well-known in classic HPC, their application to HPC-QC integration has either just begun to be explored~\cite{osti_2006942} or been overlooked in the literature, to the best of our knowledge. We believe their use, individually or in combination, will enable effective HPC-QC integration, even in the NISQ era.

\begin{figure}[t]
  \centering
  \includegraphics[width=0.75\linewidth]{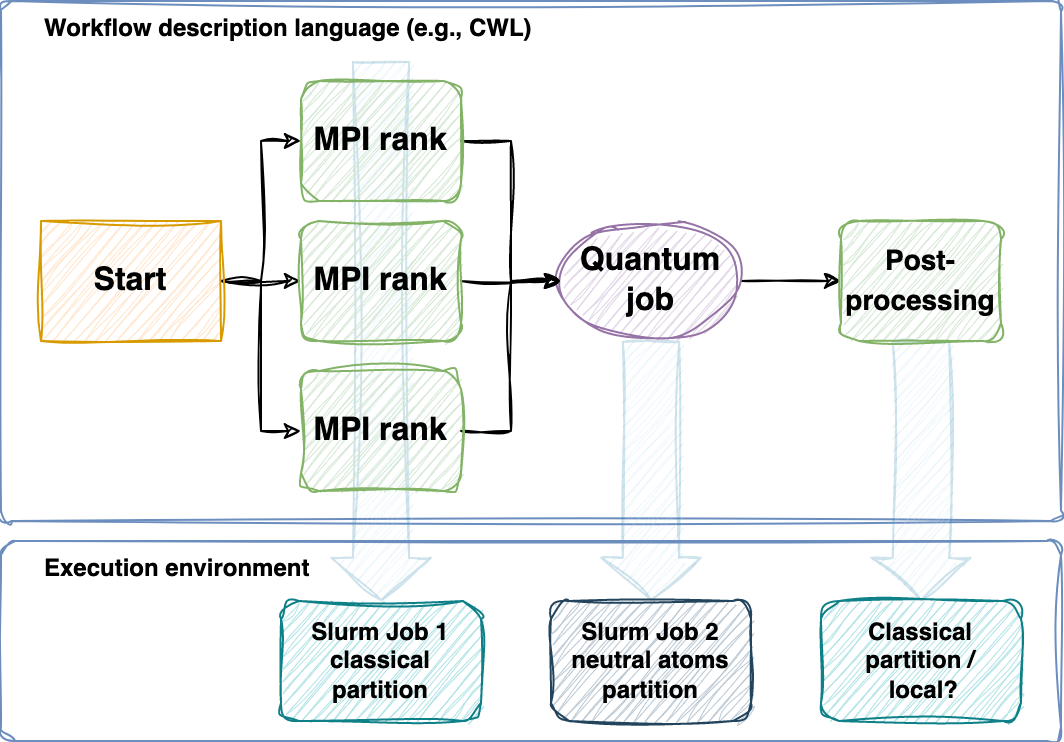}
  \caption{A Quantum-HPC workload executed as a loosely-coupled, independently scheduled workflow.}\label{fig:workflow}
\end{figure}

\noindent\textbf{Workflows}:
Loosely-coupled, \emph{workflow-based} approach leveraging well-known workflow managers (like Nextflow~\cite{di2017nextflow}, StreamFlow~\cite{StreamFlow}, PyCOMPSs~\cite{pycompss}, or even directly shell scripts) allow users to schedule classical and quantum jobs independently, but treating them as a single execution, as shown in Fig.~\ref{fig:workflow}. 
Unlike a pure co-scheduling discussed in the previous section, resource allocation does not happen initially for both quantum and classical resources; it occurs as execution requires the resources. This behaviour ensures that the execution only holds the resources it uses during the computation. On the other hand, the queuing time that each step has to go through may introduce a significant overhead when its duration outweighs the length of the computation.


\noindent\textbf{Virtual QPUs}:
As QPUs are scarce, it seems logical to introduce some mechanism to allow multiple applications to share it. While dividing the available qubits among the applications is unfeasible due to isolation issues, making them share the resource through temporal interleaving is reasonable. This statement is especially true when applications spend much time performing long running classical computation interleaved with very short quantum jobs. In this context, a pure co-scheduling solution would reserve the quantum resource without using it for too long, while a workflow scenario would suffer from significant queue times compared to the short timings of the quantum shots.

Temporal interleaving of quantum executions can be realised through virtualisation: the definition of a (fixed or dynamic) number of \emph{virtual QPUs} (VQPUs), through the configuration of a number of gres of type QPU ({\texttt{--gres=qpu}) allocated for a single real QPU. Through virtualisation, applications can request virtual QPUs, while in reality, they are reserving a time share for the execution on the same physical QPU. Fig.~\ref{fig:vqpu} shows this behaviour in action: the two applications are issuing requests for the quantum resource simultaneously, but since the rate of these requests is limited, we can co-schedule them on the resource with minimal delays, bounded by the number of VQPUs. Using VQPUs provides several benefits, as these changes do not affect the application code at all while enlarging quantum resource availability. On the other hand, if the time needed by the quantum partition is comparable to or greater than the one required to prepare the data for the shots, performing time interleaving should result in marginal gains and be unfeasible. 

\begin{figure}[t]
  \centering
  \includegraphics[width=0.75\linewidth]{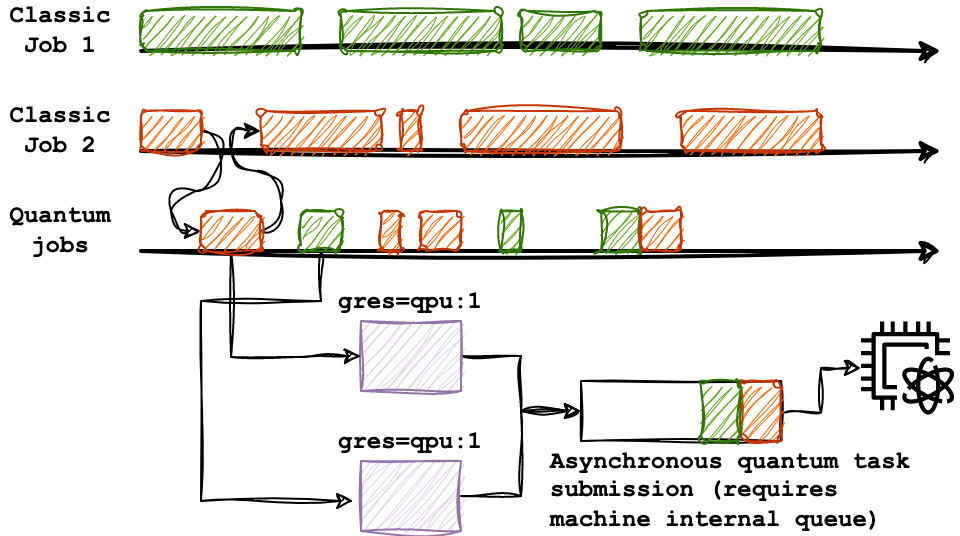}
  \caption{Representation of the asynchronous job submission to a real QPU leveraging virtual QPUs to allow multitenancy.}\label{fig:vqpu}
\end{figure}

\noindent\textbf{Malleability}:
The challenge of allocating resources for hybrid HPC-QC workloads arises from a mismatch between the dynamic nature of resource demands and the static nature of traditional resource allocation. A malleability-based approach addresses this by allowing applications to request or release resources during execution, improving the utilisation of both classical and quantum resources. Malleability has been widely explored in the HPC literature~\cite{huang2003adaptive,martin2015enhancing,iserte2020dmrlib,tarraf2024malleability} due to its significant potential benefits. However, it also introduces added complexity: applications must be capable of adapting to runtime changes in the number of available processes. This requirement has limited its adoption, as most HPC libraries, such as MPI, and the applications built on them do not natively support malleability. Nonetheless, in the emerging HPC-QC landscape, where hybrid applications are just beginning to take shape, there is an opportunity to design them with malleability in mind from the start.
\begin{figure}[t]
  \centering
  \includegraphics[width=0.9\linewidth]{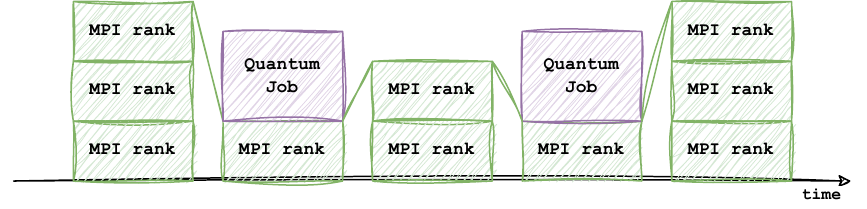}
  \caption{Representation of an use case featuring malleability properties to handle the release and retrieval of resources when switching from classical to Quantum-oriented workloads.}\label{fig:malleable}
\end{figure}

Through a malleable execution, the application dynamically requests resources as needed. Unlike workflows, resource allocation can vary at runtime based on system availability: the application can continue with fewer resources, accepting slower performance in exchange for reduced queue times. Importantly, the execution is treated as a single job rather than a sequence of tasks, avoiding repeated queuing. For instance, during the quantum phase, the job can retain minimal classical resources, enabling a faster resumption of classical computation afterward (see Fig.~\ref{fig:malleable}). This approach is especially advantageous when both the classical and quantum phases are short, as workflows may suffer from excessive queuing overhead. It can also be effective in scenarios where virtualisation yields limited benefits, since malleability is independent of quantum phase duration. However, unlike the other strategies, implementing malleability requires significant modifications to application code, as it must support dynamic adaptation to resource changes during runtime.

\section{Summary and future work}\label{sec:conclusion}


The growing interest in using QCs as accelerators for HPC clusters stems from their overlapping application domains and the complementary strengths of the two technologies. Quantum HW vendors and supercomputing centres alike are working toward practical, near-term HPC–quantum integration. However, the current landscape marked by QPU scarcity, heterogeneous technologies, and a client-server access model, demands tailored solutions for efficient resource allocation and scheduling. Simply attaching a QPU to an HPC scheduler risks underutilising either the quantum or classical resources. Depending on the workload imbalance, often influenced by the specific quantum technology, dedicated strategies are required to maximise system efficiency.

In this work, we examine this challenge and propose three complementary ideas: loosely coupled workflows, virtual QPUs, and job malleability. Each targets different scenarios and HW architectures. Future work will expand on these concepts to develop an open, practical blueprint for integrating quantum and HPC systems.


\bibliographystyle{IEEEtran}
\bibliography{references.bib}

\end{document}